\input{aipcheck}

\documentclass[
    ,final            % use final for the camera ready runs
  ]
  {aipproc}

\layoutstyle{6x9}

\begin{document}

\title{GPD and PDF modeling in terms of effective light-cone wave functions}

\classification{11.30.Cp, 12.39.Ki, 13.40.Gp, 13.60.Fz}
\keywords      {GPDs, unintegrated PDFs, overlap representation of LCWFs, spectator quark model}

%\keywords      {generalized parton distributions,
%transverse momentum dependent parton distributions,
%overlap representation,  spectator quark model
%}

\author{Dieter Mueller\footnote{
Talk given by D.M. at the XIX International Workshop on Deep-Inelastic Scattering and Related Subjects, April 11-15, 2011, 
Newport News, USA. 
} }{
  address={Nuclear Science Division, Lawrence Berkeley National Laboratory,
 Berkeley, CA 94720}
  }

\author{Dae Sung Hwang}{
  address={Department of Physics, Sejong
University, Seoul 143--747, South Korea}
}

\begin{abstract}
We employ models from effective two-body light-cone wave functions (LCWFs) to provide a link between generalized parton distributions (GPDs) and
unintegrated parton distribution functions (uPDFs).  Since we utilize the underlying Lorentz symmetry, GPDs can be entirely obtained from the parton number conserved LCWF overlaps. This also allows us to derive model constraints among GPDs. We illustrate that transversity distributions may be rather sizeable.
\end{abstract}

\maketitle

%%%%%%%%%%%%%%%%%%%%%%%%%%%%%%%%%%%%%%%%%%%%
%% MAINMATTER
%%%%%%%%%%%%%%%%%%%%%%%%%%%%%%%%%%%%%%%%%%%%

\section{Introduction}

In the field theoretical study of the hadron structure the light-cone Hamiltonian approach \cite{DreLevYan69,BroDre80,BroPauPin97} is rather
inspiring, since it describes a hadron in terms of partonic, i.e., quark and gluon, degrees of freedom.
Thereby, the LCWFs $\psi^S_{n}(X_i,{\bf k}_{\perp
i},s_i)$ are the probability amplitudes for their
corresponding $n$-parton states $|n, p^+_i, {\bf k}_{\perp
i},s_i\rangle$, which build up the proton state with spin projection $S=\{+1/2 (\Rightarrow),-1/2(\Leftarrow)\}$ along the $z$-axis:
\begin{eqnarray}
\label{Def-ProSta} |P,S\rangle = \sum_n
\int\![dX\, d^2{\bf k}_\perp]_n\, \prod_{j=1}^n \frac{1}{\sqrt{X_j}}\; \psi^S_n(X_i,{\bf
k}_{\perp i},s_i)\, |n, X_i
P^+, X_i {\bf P}_\perp + {\bf k}_{\perp i},s_i\rangle\,.
\end{eqnarray}
Here, the parton $i$ has longitudinal momentum fraction $X_i$, transverse momentum $X_i {\bf P}_\perp +{\bf k}_{\perp i}$,  spin projection  $s_i$, and  $[dX\, d^2{\bf k}_\perp]_n$ denotes the $n$-parton phase space, cf.~\cite{HwaMue07}.
The LCWFs have been used for the investigation of the hadron properties \cite{BroPauPin97} and the LCWF overlap representation of GPDs was studied in \cite{BroDieHwa00,DieFelJakKro00,Mukherjee:2002gb,HwaMue07}.

When we consider a scalar--diquark spectator model, we have only four LCWFs \cite{HwaMue07},
\begin{eqnarray}
\label{Def-LC-WF1} &&\!\!\!\!\!\!\!\Psi^{\Rightarrow}_{\rightarrow}%(X,{\bf k}_\perp|\lambda^2)
=\frac{m+X M}{M \sqrt{1-X}}\sqrt{\rho(\lambda)}\,
\phi(X,{\bf k}_\perp|\lambda),\quad
\Psi^{\Rightarrow}_{\leftarrow}%(X,{\bf k}_\perp|\lambda^2)
= \frac{-k^1- i k^2}{M\sqrt{1-X}}\sqrt{\rho(\lambda)}\, \phi(X,{\bf k}_\perp|\lambda),
\\
\label{Def-LC-WF2} &&\!\!\!\!\!\!\!
\Psi^{\Leftarrow}_{\leftarrow}%(X,{\bf k}_\perp|\lambda^2)
=\frac{m+X M}{M\sqrt{1-X}}\sqrt{\rho(\lambda)}\,
\phi(X,{\bf k}_\perp|\lambda), \quad
\Psi^{\Leftarrow}_{\rightarrow}%(X,{\bf k}_\perp)
 = \frac{k^1- i k^2}{M\sqrt{1-X}}\sqrt{\rho(\lambda)}\, \phi(X,{\bf k}_\perp|\lambda),
\end{eqnarray}
in terms of one {\em effective} LCWF $\phi(X,{\bf k}_\perp|\lambda)$, which we weight with the square root of the spectral density  $\rho(\lambda)$.
Here, $\lambda$ is the scalar diquark spectator mass, $m$ is the struck quark mass, and  $M$ is the proton mass.   In (\ref{Def-LC-WF1}--\ref{Def-LC-WF2}) the proton has its spin projection $(\Rightarrow)$ or $(\Leftarrow)$ along the $z$-axis, and the constituent quark state has the aligned ($\rightarrow$) or opposite ($\leftarrow$) spin projection and the orbital angular momentum projection $L^z=0$ or $L^z=\pm 1$.

\section{LCWF overlap and GPD constraints}

We define in our model the scalar LCWF overlap in the outer region $x \ge \eta$ as
\begin{eqnarray}
\label{GPD-overlap}
\mbox{\boldmath $\Phi$}(x \ge \eta,\eta,\mbox{\boldmath $\Delta$}_{\perp},{\bf
k}_{\perp}) =
\int\!\! d\lambda^2\; \frac{\rho(\lambda^2)}{1-x}\,
\phi^\ast\!\!\left(\!\!\frac{x-\eta}{1-\eta},{\bf k}_{\perp}- \frac{1-x}{1-\eta} \mbox{\boldmath $\Delta$}_\perp\bigg|\lambda\!\!\right)
\phi\!\!\left(\!\!\frac{x+\eta}{1+\eta},{\bf k}_{\perp}\bigg|\lambda\!\!\right),
\end{eqnarray}
which represents a unintegrated GPD (uGPD). In the forward limit, i.e.,
$
\mbox{\boldmath $\Phi$}(x,{\bf k}_\perp)\equiv \mbox{\boldmath $\Phi$}(x,\eta=0,\mbox{\boldmath $\Delta$}_{\perp}=0,{\bf
k}_{\perp}) \ ,
$
it reduces to a uPDF, and the integration over ${\bf k}_\perp$ gives a GPD or a PDF.
Since various partonic quantities can be expressed in terms of the LCWF overlap (\ref{GPD-overlap}), we have their model dependent relations, for example, the relations among the uPDFs which were extensively studied, see, e.g., Ref.~\cite{Avaetal09} and
references therein.

To construct the GPD in the central region, i.e., $|x|\le \eta$, it is useful to employ the double distribution (DD) representation \cite{MueRobGeyDitHor94,Rad97}, which reads in the unintegrated form as
\begin{eqnarray}
\label{Def-DDunint}
\mbox{\boldmath $\Phi$}(x,\eta,\mbox{\boldmath $\Delta$}_{\perp},{\bf k}_{\perp}) &\!\!\!=\!\!\!& \int_0^1\! dy\int_{-1+y}^{1-y}\! dz\;  \delta(x-y-z \eta)\,
 \int\!\! d\lambda^2\; \rho(\lambda^2)\, \mbox{\boldmath $\hat\Phi$}\!\left(y,z,t,\overline{\bf k}_\perp\big| \lambda\!\right),
\end{eqnarray}
where $\overline{{\bf k}}_\perp = {\bf k}_\perp -(1-y+z) \mbox{\boldmath $\Delta$}_\perp/2$ and $ \mbox{\boldmath $\hat\Phi$}$ is the Laplace transform of the LCWF overlap,
\begin{eqnarray}
\label{Def-DD-unint1}
&&\!\!\!\!\!\! \mbox{\boldmath $\hat\Phi$}(y,z,t,\overline{{\bf k}}_\perp\big| \lambda)= \frac{1}{2}\int_0^\infty\!dA\,A\;
\varphi^\ast\!\left(\!A \frac{1-y+z}{2}\!\right) \varphi\!\left(\!A \frac{1-y-z}{2}\!\right)
\\
&&\times
\exp \left\{-A\left[
(1-y)\frac{m^2}{M^2}+y \frac{\lambda^2}{M^2}-y(1-y)-\left[(1-y)^2-z^2\right] \frac{t}{4 M^2}+\frac{\overline{{\bf k}}^2_\perp}{M^2}
\right] \right\},\nonumber
\end{eqnarray}
which depends besides the mass parameters also on the functional form of the reduced LCWF $\varphi(\alpha)$.
Note that this representation is inspired from a commonly used spectator quark model, which can be expressed in terms of a triangle diagram.

The eight twist-two GPDs $F\in \{H,E,\widetilde H, \widetilde E, H_{\rm T},E_{\rm T},\widetilde H_{\rm T}, \widetilde E_{\rm T}\}$ can be straightforwardly expressed in terms of the scalar LCWF overlap \cite{HwaMue07}, e.g., in the chiral odd sector:
\begin{eqnarray}
F_{\rm T}(x,\eta,t) =
\int_{0}^1\!dy\; \int_{-1+y}^{1-y}\! dz\;
\delta(x-y-\eta z ) f_{\rm T}(x,\eta,t)\,,
\end{eqnarray}
where the DDs are given in terms of the scalar LCWF overlap (\ref{Def-DDunint}), integrated over ${\bf k}_\perp$:
\begin{eqnarray}
\label{hT-thT}
\!\!\!\!\!\!\!\!\! h_{\rm T}=
\left[\!
\left(\!\frac{m}{M}+y\!\right)^2-\left((1-y)^2-z^2\right)\frac{t}{4M^2}\! \right] \mbox{\boldmath $\hat\Phi$}(y,z,t),
&&\!\!\!\!\!\!\!\!\!
\widetilde h_{\rm T}  =
-\left[(1-y)^2-z^2\right]\mbox{\boldmath $\hat\Phi$}(y,z,t),
\\
\label{eT-teT}
\!\!\!\!\!\!\!\!\! e_{\rm T}=
2\left[\!\left(\!\frac{m}{M}+y\!\right)(1-y)+(1-y)^2-z^2\!\right]\mbox{\boldmath $\hat\Phi$}(y,z,t),
%\\
%\label{Def-SpeFun-teT}
&&\!\!\!\!\!\!\!\!\!
\widetilde e_{\rm T} =
%&\!\!\!=\!\!\! &
2\left(\frac{m}{M}+y\right) z\, \mbox{\boldmath $\hat\Phi$}(y,z,t).
\end{eqnarray}
The factor $(m/M+y)^2$  in $h_{\rm T}$ arises from the diagonal $L^z=0$ LCWF overlap, while the $t$-dependent part comes from an
overlap of $L^z=+1$ and $L^z=-1$ LCWFs ($\Delta L^z=2$).
The ${\widetilde h}_T$  DD  is  entirely induced from a $\Delta L^z=2$ LCWF overlap
and it is related to the so-called "pretzelosity" uPDF. The ${\widetilde e}_T$ DD contains a $|L^z|=1$  and $L^z=0$  overlap ($\Delta L^z=1$)
and ${e}_T$ DD contains both $\Delta L^z=1$ and $\Delta L^z=2$ overlaps.

Obviously, we can write down various relations among different DDs and express the chiral odd ones by the chiral even ones.
Let us consider here only the chiral odd GPDs
\begin{equation}
\label{ET}
\overline{H}_T = H_T - \frac{t}{4M^2} \widetilde{H}_T\,,\quad
\overline{E}_T = E_T + 2 \widetilde{H}_T\,,
\end{equation}
in which the $\Delta L^z =2$ overlap is removed.
In any scalar diquark model these chiral odd GPDs differ from the chiral even ones by a  $\eta$ proportional term,
which is given by  $\widetilde{E}_{\rm T}$:
\begin{eqnarray}
\label{bH-bE}
\overline{H}_T(x,\eta,t) \stackrel{\rm mod}{=} \frac{1}{2} \left[H+ \widetilde H -\eta \widetilde{E}_{\rm T}\right](x,\eta,t),
\;\;
\overline{E}_{\rm T}(x,\eta,t)\stackrel{\rm mod}{=}E(x,\eta,t) + \eta \widetilde{E}_{\rm T}(x,\eta,t).
\end{eqnarray}
The chiral odd GPD $\widetilde{E}_{\rm T}$ vanishes by itself in the zero-skewness case, since the corresponding DD
(\ref{eT-teT}) is odd in $z$. Hence, for small or moderate $\eta$  values both
GPDs (\ref{bH-bE}) are approximately given by the corresponding chiral even GPDs.

\section{Model predictions for chiral odd GPDs}

\begin{figure}[t]
\includegraphics[height=.235\textheight]{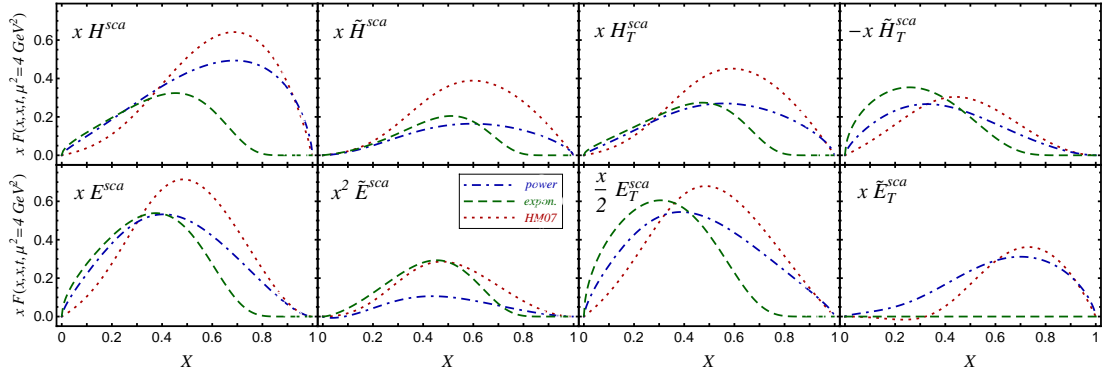}
\caption{\label{Fig-GPDcom1} \small
Our Regge improved power-like (dash-dotted), exponential (dashed) LCWF, and {\it HM07} (dotted) \cite{HwaMue07} models are
displayed for valence quark  GPDs $F^{\rm sca}=2 F^{u_{\rm val}}/3- F^{d_{\rm val}}/3$ at $\eta =x$ and $t=-0.2\, {\rm GeV}^2$ versus $X=2x/(1+x)$.
Upper [lower] panels from left to right: $x H$ [$x E$] ,  $\widetilde H$ [$x^2 \widetilde E$], $H_{\rm T}$ [$x E_{\rm T}/2$], and
$-\widetilde H_{\rm T}$ [$x {\widetilde E}_{\rm T}$].}
\end{figure}
Assuming SU(6) symmetry, let us now provide some GPD predictions for the scalar diquark contents, which have the quark combination $2u/3  - d/3$. Taking the functional form  of the reduced LCWF, which enters in the unintegrated DD (\ref{Def-DD-unint1}), as
 \begin{eqnarray}
\label{HMpow-varphi}
\varphi^{\rm pow.}(\overline{\alpha}) = \frac{g\, \overline{\alpha}^{p+\alpha/2}}{M \Gamma(1+p+\alpha/2)}\,,
\qquad
\varphi^{\rm exp.}(\overline{\alpha}) = \frac{g\,}{M} \delta(\overline{\alpha} - \bar{A})\,,
\end{eqnarray}
and adopting a Regge inspired form for the spectral density \cite{LanPol71}
\begin{eqnarray}
\label{Def-SpeRep1}
\rho_\alpha(\lambda,\lambda_{\rm cut})  =
\theta(\lambda^2-\lambda_{\rm cut}^2)  \frac{(\lambda^2 -\lambda_{\rm cut}^2)^{\alpha-1}}{\Gamma(\alpha)  M^{2\alpha} }\,,
\end{eqnarray}
we build three models, two with power-like  and one with exponential ${\bf k}_\perp$ fall-off. Here, $g$ is the coupling of the struck quark to the spectator system, $p$ determines the power-like fall-off of ${\bf k}_\perp$-dependence, $\alpha$ is taken
to be close to the $\rho$ Regge trajectory or, alternatively, set to zero, and $\overline{A}$ determines the slope of the exponential ${\bf k}_\perp$-dependence.
All these parameters together with the struck quark mass and the spectator threshold $\lambda_{\rm cut}$ are fitted to the functional shape of PDFs and form factors, where our Regge improved power-like model provides a good agreement. The GPD predictions for the eight twist-two GPDs at the cross-over line $\eta=x$ are shown in Fig.~\ref{Fig-GPDcom1}. Note that this quantity determines at leading order the imaginary
part of hard exclusive amplitudes, where the real part can be obtained from a partonic dispersion relation. As one realizes, the chiral odd quantities, including the "pretzelosity" related $\widetilde H_{\rm T}$ GPD, are rather sizeable.

\section{Conclusions}

Utilizing the underlying Lorentz symmetry through the DD, parton number conserved LCWF overlap representations are employed to model GPDs in the entire momentum fraction region. In a scalar diquark model one finds easily relations among GPDs which are the analogies to the relations among uPDFs. Moreover, the unintegrated
DD given in (\ref{Def-DD-unint1}) tells us that in such models $t$-dependency in GPDs and ${\bf k}_\perp$-dependency in uPDFs are tied to each other, however, a one-to-one correspondence
can be obtained only when the DD is concentrated at one specific $z$ value, e.g., $z=0$. Finally,
effective LCWF overlap representations may become a useful phenomenological tool to quantify the partonic contents in a global fitting procedure to both exclusive and inclusive observables.

%%%%%%%%%%%%%%%%%%%%%%%%%%%%%%%%%%%%%%%%%%%%
%% Sample figure:
%%
%% The option [height=...] scales the picture to the given height,
%% without it it would be printed at its nominal size
%%%%%%%%%%%%%%%%%%%%%%%%%%%%%%%%%%%%%%%%%%%%

%%%%%%%%%%%%%%%%%%%%%%%%%%%%%%%%%%%%%%%%%%%%%%%%
%% BACKMATTER
%%%%%%%%%%%%%%%%%%%%%%%%%%%%%%%%%%%%%%%%%%%%%%%%

\begin{theacknowledgments}
DM is indebted to the members of the Nuclear Science Division at the
Lawrence Berkeley National Laboratory for the warm hospitality
during his visit. He is grateful to H.~Avakian, P.~Schweitzer, and  F.~Yuan for
stimulating discussions.
This work was supported in part by
Korea Foundation for International Cooperation of Science \& Technology (KICOS)
and National Research Foundation of Korea (2011-0005226).
\end{theacknowledgments}



\end{document}